\documentclass[aps,prl,preprint,superscriptaddress,11pt]{revtex4}

\usepackage{amsmath}
\usepackage{graphics}
\usepackage{graphicx}
\usepackage[mathcal]{eucal}

\begin{document}

\preprint{COLO-HEP-545, UCI-TR-2009-9}

\title{Group Theoretical Origin of CP Violation}

\author{Mu-Chun Chen}
\email[]{muchunc@uci.edu}
\affiliation{Department of Physics \& Astronomy, 
University of California, Irvine, CA 92697-4575, USA}
\author{K.T. Mahanthappa}
\email[]{ktm@pizero.colorado.edu}
\affiliation{Department of Physics, University of Colorado at Boulder, Boulder, CO 80309-0390, USA}


\begin{abstract}
We propose the complex group theoretical Clebsch-Gordon coefficients as a novel origin of CP violation. This is manifest in our model based on SU(5)  combined with the 
$T^{\prime}$ group as the family symmetry. 
The complex CG coefficients in $T^{\prime}$ lead to explicit CP violation which is thus geometrical in origin. 
The predicted CP violation measures in the quark sector are consistent with the current experimental data. 
The corrections due to leptonic Dirac CP violating phase gives the experimental best fit value for the solar mixing angle, and we also gets the right amount of the baryonic asymmetry. 
\end{abstract}

\pacs{}

\maketitle


The origin of the cosmological matter antimatter asymmetry in the universe is one of the fundamental questions that still remain to be answered. It has long been known that in order to generate the baryonic asymmetry, three conditions~\cite{Sakharov:1967dj} must be satisfied, {\it i.e.} baryon and lepton number violations, CP violation and out-of-equilibrium decay. Given the evidence that  our universe is expanding, the out-of-equilibrium condition can be simply satisfied. In most extensions of the Standard Model, such as grand unified theories, there naturally exist processes that violate baryon and/or lepton numbers. Due to the small quark mixing, the complex phase in the Cabibbo-Kobayashi-Maskawa (CKM) mixing matrix generates a baryonic asymmetry that is vanishingly small. The observation of neutrino oscillation, on the other hand, opens up the possibility of generating the baryonic asymmetry through leptogenesis~\cite{Fukugita:1986hr}. The success of leptogenesis crucially depends on the existence of CP violating phases in the Pontecorvo-Maki-Nakagawa-Sakata (PMNS) matrix that describes the neutrino mixing~\cite{Chen:2007fv}.

Generally, CP violation can arise either explicitly through complex Yukawa coupling constants, or spontaneously through the complex vacuum expectation values (VEVs) of the Higgs fields, or a combination of both. In these two scenarios, the complex phases appear to be free parameters, adding to the list of parameters in the Yukawa sector that accommodate the observed fermion masses, mixing angles and CP violation measures.  

In this letter, we propose the complex Clebsch-Gordon (CG) coefficients as a new origin of CP violation. Such complex CG coefficients exist in the double tetrahedral group, $T^{\prime}$. In this scenario, CP violation occurs explicitly in the Lagrangian from the CG coefficients of the $T^{\prime}$ group theory, while the Yukawa coupling constants and the VEVs of the scalar fields remain real. As a result, the amount of CP violation in our model is determined entirely by the group theory, unlike in the usual scenarios. It gives the right amount of CP violation in the quark sector and in the lepton sector, through leptogenesis, gives rise to the right amount of the matter antimatter asymmetry.

Experimentally, the best fit values for the neutrino mixing angles are very close to the prediction of the tri-bimaximal mixing (TBM) matrix~\cite{Harrison:2002er},
\begin{equation}
U_{TBM} = \left(\begin{array}{ccc}
\sqrt{2/3} & \sqrt{1/3} & 0 \\
-\sqrt{1/6} & \sqrt{1/3} & -\sqrt{1/2} \\
-\sqrt{1/6} & \sqrt{1/3} & \sqrt{1/2}
\end{array}\right) \; 
\end{equation}
which predicts
$\sin^{2} \theta_{atm} = 1/2$, $\tan^{2}\theta_{\odot} = 1/2$ and $\sin\theta_{13} = 0$. 
It has been realized that the TBM matrix can arise from an underlying $A_{4}$ symmetry~\cite{Ma:2001dn}. Nevertheless, $A_{4}$ does not give rise to quark mixing~\cite{Ma:2006sk}. Even though the exact TBM matrix does not give rise to CP violation, due to the correction from the charged lepton sector in our model, 
leptonic CP violation can still arise. 

{\it Group Theory of $T^{\prime}$.}---The finite group $T^{\prime}$ is the double covering group of the tetrahedral group, $A_{4}$. It has 24 elements, and is generated by two generators, $S$ and $T$. 
In the $T$ diagonal basis,  these two generators in the triplet representation are given by,
\begin{equation}\label{eq:gen}
S = \frac{1}{3} \left(\begin{array}{ccc}
-1 & 2\omega & 2\omega^{2} \\
2\omega^{2} & -1 & 2\omega \\
2\omega & 2\omega^{2} & -1
\end{array}\right) \; , \; 
T = \left(\begin{array}{ccc}
1 & 0 & 0 \\
0 & \omega & 0 \\
0 & 0 & \omega^{2}
\end{array}\right) \; ,
\end{equation}
with $\omega = e^{2i\pi/3}$. 
While all CG coefficients can be chosen to be real in $A_{4}$, this is not the case in $T^{\prime}$, which has three doublet representations, $2, \; 2^{\prime}, \; 2^{\prime\prime}$, in addition to the triplet, $3$, and three singlet representations, $1, \; 1^{\prime}, \; 1^{\prime\prime}$, that exist in $A_{4}$.  
Specifically, the complex CG coefficients appear in the products that involve the doublet representations~\cite{cg}.

{\it The Model.}---In Ref.~\cite{Chen:2007afa}, we have constructed a $SU(5)$ model combined with a family symmetry based on $T^{\prime}$, which  simultaneously gives rise to the tri-bimaximal neutrino mixing and realistic CKM quark mixing~\cite{Chen:2003zv}. 
($T^{\prime}$ has also been utilized by others~\cite{Feruglio:2007uu}.) In this investigation~\cite{Chen:2007afa}, we did not consider the possibility of having CP violation due to CG coefficients being complex, which is the subject matter of the present paper. 
The field content of our model is summarized in Table~\ref{tbl:charge}. Note that since all fields in a full $SU(5)$ multiplet transform in the same way under the $T^{\prime}$ symmetry, our model is free of discrete gauge anomalies automatically~\cite{Luhn:2008xh,Chen:2006hn}. In addition to the $SU(5)\times T^{\prime}$ symmetry, we further impose a $Z_{12} \times Z_{12}^{\prime}$ symmetry. Due to the $Z_{12} \times Z_{12}^{\prime}$ symmetry, only nine operators are allowed in our model up to mass dimension-7 in the Yukawa sector. The discrete symmetries of our model allow the lighter generation masses to arise only at higher mass dimensionality, and thus provides a dynamical origin of the mass hierarchy. 

The Lagrangian of the Yukawa sector of the model is given by,  
$\mathcal{L}_{\mbox{\tiny Yuk}} =  \mathcal{L}_{TT} + \mathcal{L}_{TF} + \mathcal{L}_{FF} + h.c.$, where  
\begin{eqnarray}
-\mathcal{L}_{TT} & = & y_{t} H_{5} T_{3} T_{3} + \frac{1}{\Lambda^{2}}  H_{5} \biggl[ y_{ts} T_{3} T_{a} \psi \zeta + y_{c} T_{a} T_{b} \phi^{2} \biggr] + \frac{1}{\Lambda^{3}} y_{u} H_{5} T_{a} T_{b} \phi^{\prime 3} \; , \label{eq:Ltt} \\ 
-\mathcal{L}_{TF} & = &  \frac{1}{\Lambda^{2}} y_{b} H_{\overline{5}}^{\prime} \overline{F} T_{3} \phi \zeta + \frac{1}{\Lambda^{3}} \biggl[ y_{s} \Delta_{45} \overline{F} T_{a} \phi \psi N  + y_{d} H_{\overline{5}^{\prime}} \overline{F} T_{a} \phi^{2} \psi^{\prime} \biggr] \; ,    \label{eq:Ltf} \\
-\mathcal{L}_{FF} & = & \frac{1}{\Lambda M_{X}} \biggl[ \lambda_{1} H_{5} H_{5} \overline{F}\overline{F} \xi + \lambda_{2} H_{5} H_{5} \overline{F}\overline{F} \eta\biggr] \; ,
\label{eq:Lff}
\end{eqnarray}
which is invariant under $SU(5) \times T^{\prime}$ and it is CP non-invariant. 
Here the parameter $\Lambda$ is the cutoff scale of the $T^{\prime}$ symmetry while $M_{X}$ is the scale where lepton number violating operators are generated. Note that all Yukawa coupling constants, $y_{x}$, in the Lagrangian are real parameters. Even if they are made complex, their phases can be absorbed by redefinition of the Higgs and flavon fields. 
The $T^{\prime}$ flavon fields acquire vacuum expectation values along the following direction,
\begin{eqnarray}
\left<\xi\right> = \left(\begin{array}{c}
1 \\ 1 \\ 1
\end{array}\right)
\xi_{0} \Lambda \; , \; 
\left< \phi^{\prime} \right> = \left(\begin{array}{c}
1 \\ 1 \\ 1
\end{array}\right) \phi_{0}^{\prime} \Lambda \; , \;  \\
\left< \phi \right> = \left( \begin{array}{c} 
0 \\ 0 \\ 1
\end{array}\right) \phi_{0} \Lambda \; , \; 
\left< \psi \right> = \left( \begin{array}{c} 1 \\ 0 \end{array}\right)
\psi_{0} \Lambda \; , \; \; \\
\left< \psi^{\prime} \right> = \left(\begin{array}{c} 1 \\ 1 \end{array}\right) \psi_{0}^{\prime} \Lambda \; , \\
\left< \zeta \right> = \zeta_{0} \Lambda \; , \; \left< N \right> = N_{0} \Lambda \; , \; \left< \eta \right> = u_{0} \Lambda \; .
\end{eqnarray}
Note that  all the expectation values are real and they don't contribute to CP violation.  
(An interesting possibility of having spontaneous CP violation  even though the VEVs of scalars are real has been discussed~\cite{Masiero:1998yi}.) 
\begin{table}
\begin{tabular}{|c|ccc|ccc|cccccc|cc|}\hline
& $T_{3}$ & $T_{a}$ & $\overline{F}$ & $H_{5}$ & $H_{\overline{5}}^{\prime}$ & $\Delta_{45}$ & $\phi$ & $\phi^{\prime}$ & $\psi$ & $\psi^{\prime}$ & $\zeta$ & $N$ & $\xi$ & $\eta$  \\ [0.3em] \hline\hline
SU(5) & 10 & 10 & $\overline{5}$ & 5 &  $\overline{5}$ & 45 & 1 & 1 & 1 & 1& 1 & 1 & 1 & 1\\ \hline
$T^{\prime}$ & 1 & $2$ & 3 & 1 & 1 & $1^{\prime}$ & 3 & 3 & $2^{\prime}$ & $2$ & $1^{\prime\prime}$ & $1^{\prime}$ & 3 & 1 \\ [0.2em] \hline
$Z_{12}$ & $\omega^{5}$ & $\omega^{2}$ & $\omega^{5}$ & $\omega^{2}$ & $\omega^{2}$ & $\omega^{5}$ & $\omega^{3}$ & $\omega^{2}$ & $\omega^{6}$ & $\omega^{9}$ & $\omega^{9}$ 
& $\omega^{3}$ & $\omega^{10}$ & $\omega^{10}$ \\ [0.2em] \hline
$Z_{12}^{\prime}$ & $\omega$ & $\omega^{4}$ & $\omega^{8}$ & $\omega^{10}$ & $\omega^{10}$ & $\omega^{3}$ & $\omega^{3}$ & $\omega^{6}$ & $\omega^{7}$ & $\omega^{8}$ & $\omega^{2}$ & $\omega^{11}$ & 1 & $1$ 
\\ \hline   
\end{tabular}
\vspace{-0.in}
\caption{Field content of our model. The three generations of matter fields in $10$ and $\overline{5}$ of $SU(5)$ are in the $T_{3}$, $T_{a}$ $(a=1,2)$ and $\overline{F}$ multiplets. The Higges that are needed to generate $SU(5)$ invariant Yukawa interactions are $H_{5}$, $H_{\overline{5}}^{\prime}$ and $\Delta_{45}$. The flavon fields $\phi$ through $N$ are those that give rise to the charged fermion mass matrices, while $\xi$ and $\eta$ are the ones that generate neutrino masses.  The $Z_{12}$ charges are given in terms of the parameter $\omega = e^{i\pi/6}$.}  
\label{tbl:charge}
\end{table}

In terms of the $T^{\prime}$ and $SU(5)$ component fields, the above Lagrangian gives the following Yukawa interactions for the charged fermions in the weak charged current interaction eigenstates,
\begin{eqnarray}\label{eq:L2}
-\mathcal{L}_{\mbox{\tiny Yuk}} &  \supset &  \overline{U}_{R, i} (M_{u})_{ij} Q_{L,j} + \overline{D}_{R,i} (M_{d})_{ij} Q_{L,j}  + \overline{E}_{R,i} (M_{e})_{ij} \ell_{L,j} + h.c. \; ,
\end{eqnarray}
where $Q_{L}$ denotes the quark doublets while $U_{R}$ and $D_{R}$ denotes the iso-singet up- and down-type quarks, with $i$ and $j$ being the generation indices. Similarly, $\ell_{L}$ and $E_{R}$ denote the iso-doublet and singlet charged leptons, respectively. The matrices $M_{u}$, $M_{d}$ and $M_{e}$, upon the breaking of $T^{\prime}$ and the electroweak symmetry, are given in terms of seven parameters by~\cite{Chen:2007afa}
\begin{eqnarray}
M_{u} & = & \left( \begin{array}{ccc}
i \phi^{\prime 3}_{0}  & (\frac{1-i}{2}) \phi_{0}^{\prime 3} & 0 \\
(\frac{1-i}{2})  \phi_{0}^{\prime 3}  & \phi_{0}^{\prime 3} + (1 - \frac{i}{2}) \phi_{0}^{2} & y^{\prime} \psi_{0} \zeta_{0} \\
0 & y^{\prime} \psi_{0} \zeta_{0} & 1
\end{array} \right) y_{t}v_{u}, \qquad \\
M_{d}  & = & \left( \begin{array}{ccc}
0 & (1+i) \phi_{0} \psi^{\prime}_{0} & 0 \\
-(1-i) \phi_{0} \psi^{\prime}_{0} & \psi_{0} N_{0} & 0 \\
\phi_{0} \psi^{\prime}_{0} & \phi_{0} \psi^{\prime}_{0} & \zeta_{0} 
\end{array}\right) y_{d} v_{d} \phi_{0} \; , \\
M_{e} & = & \left( \begin{array}{ccc}
0 & -(1-i) \phi_{0} \psi^{\prime}_{0} & \phi_{0} \psi^{\prime}_{0} \\
(1+i) \phi_{0} \psi^{\prime}_{0} & -3 \psi_{0} N_{0} & \phi_{0} \psi^{\prime}_{0} \\
0 & 0 & \zeta_{0} 
\end{array}\right) y_{d} v_{d} \phi_{0} \; . 
\end{eqnarray}
Here we have absorbed the couplings, $y_{d}$, $y_{s}$, $y_{c}/y_{t}$ and $y_{u}/y_{t}$, by re-scaling the VEV's, $\phi_{0}$, $\psi_{0}^{\prime}$, $\psi_{0}$, and $\phi_{0}^{\prime}$, respectively. We also define $y^{\prime} = y_{ts}/\sqrt{y_{c}y_{t}}$.  
The $SU(5)$ relation, $M_{d} = M_{e}^{T}$, is manifest in the above equations, except for the factor of $-3$ in the (22) entry of $M_{e}$, due to the $SU(5)$ CG coefficient through the coupling to $\Delta_{45}$. In addition to this $-3$ factor, the Georgi-Jarlskog (GJ) relations also require $M_{e,d}$ being non-diagonal, leading to corrections to the TBM pattern~\cite{Chen:2007afa}.  Note that the complex coefficients in the above mass matrices arise {\it entirely} from the CG coefficients of the $T^{\prime}$ group theory. More precisely, these complex CG coefficients appear in couplings that involve the doublet representations of $T^{\prime}$.

The mass matrices $M_{u,d}$ are diagonalized by,
$V_{u, R}^{\dagger} M_{u} V_{u,L}  =  \mbox{diag} (m_{u}, m_{c}, m_{t})$ and 
$V_{d, R}^{\dagger} M_{d} V_{d,L}  =  \mbox{diag} (m_{d}, m_{s}, m_{b})$,
where the mass eigenvalues on the right-hand side of the equations are real and positive. 
This gives the following weak charged current interaction in the mass eigenstates of the fermions,
\begin{eqnarray}\label{eq:Lcc}
\mathcal{L}_{cc} & = & \frac{g}{2\sqrt{2}} \biggl[ W^{\mu}_{+}(\vec{x},t) J_{\mu}^{-} (\vec{x},t) 
 + W^{\mu}_{-} (\vec{x},t) J_{\mu}^{+}(\vec{x},t) \biggr] \; , \nonumber \\
J_{\mu}^{-} & = & 
(\overline{u}^{\prime}, \overline{c}^{\prime}, \overline{t}^{\prime})_{L} \gamma_{\mu} V_{CKM}
\left( \begin{array}{c}
d^{\prime} \\
s^{\prime} \\
b^{\prime}
\end{array}\right)_{L} \; . 
\end{eqnarray}
The complex mass matrices $M_{u,d}$ lead to a complex quark mixing matrix, 
$V_{CKM} = V_{u,L}^{\dagger} V_{d,L}$.

The interactions in $\mathcal{L}_{FF}$ lead to the following neutrino mass matrix, 
\begin{equation}\label{eq:fd}
M_{\nu} = \left( \begin{array}{ccc}
2\xi_{0} + u_{0} & -\xi_{0} & -\xi_{0} \\
-\xi_{0} & 2\xi_{0} & -\xi_{0} + u_{0} \\
-\xi_{0} & -\xi_{0} + u_{0} & 2\xi_{0} 
\end{array} \right) \frac{\lambda v^{2}}{M_{x}} \; ,
\end{equation}
which is parametrized by {\it two} parameters, giving the three absolute neutrino masses~\cite{Chen:2007afa} (see below). Here the coupling $\lambda_{2}/\lambda_{1}$ has been absorbed by redefining the VEV, $u_{0}$, and $\lambda = \lambda_{1}$.  As these interactions involve only the triplet representations of $T^{\prime}$, the relevant product rule is $3 \otimes 3$. Consequently, all CG coefficients are real, leading to a real neutrino Majorana mass matrix. The neutrino mass matrix given in Eq.~\ref{eq:fd} has the special property that it is form diagonalizable~\cite{Chen:2009um}, {\it i.e.} independent of the values of $\xi_{0}$ and $u_{0}$, it is diagonalized by the tri-bimaximal mixing matrix,
$U_{\mbox{\tiny TBM}}^{T} M_{\nu} U_{\mbox{\tiny TBM}}  =    \mbox{diag}(u_{0} + 3 \xi_{0}, u_{0}, -u_{0}+3\xi_{0}) \frac{v_{u}^{2}}{M_{X}}$ 
 $\equiv   \mbox{diag} (m_{1}, m_{2}, m_{3})$.
While the neutrino mass matrix is real, the complex charged lepton mass matrix $M_{e}$, which is diagonalized by, 
$V_{e, R}^{\dagger} M_{e} V_{e,L} =  \mbox{diag} (m_{e}, m_{\mu}, m_{\tau})$, 
leads to a complex 
$V_{\mbox{\tiny PMNS}} = V_{e, L}^{\dagger} U_{\mbox{\tiny TBM}}$ (see below).

{\it CPT Invariance and CP Violation.}---Even though the complexity of the Lagrangian arises in our model through the complex CG coefficients, the hermiticity of the Lagrangian, which is required in order to have CPT invariance, remains satisfied.  This is easily seen using the component form given in Eq.~\ref{eq:L2}. Take the term $\overline{U}_{R} M_{u} Q_{L}$ for example. Its corresponding hermitian conjugate is
$(\overline{U}_{R} M_{u} Q_{L})^{\dagger} 
= (U_{R}^{\dagger} \gamma_{0} M_{u} Q_{L})^{\dagger}$ 
$= \overline{Q}_{L} M_{u}^{\dagger} U_{R}$.
The hermiticity of the Lagrangian allows us to write, in general, 
$\mathcal{L}(\vec{x},t) =  \alpha \mathcal{O}(\vec{x},t) + \alpha^{\ast} \mathcal{O}^{\dagger}(\vec{x},t)$, 
where $\mathcal{O}(\vec{x},t)$ is some operator and $\alpha$ is some c-number. 
Recall that charge conjugation $\mathcal{C}$ changes a left-handed particle into a left-handed anti-particle, while the parity $\mathcal{P}$ turns a left-handed particle into a right-handed particle, and vice versa. Thus the $\mathcal{CP}$ transformation converts a left-handed particle into a right-handed anti-particle. Effectively, 
$\mathcal{O}(\vec{x},t) \; \stackrel{\mathcal{CP}}{\longrightarrow} \; \mathcal{O}^{\dagger} (-\vec{x},t)$  and
$\alpha \stackrel{\mathcal{CP}}{\longrightarrow} \; \alpha$. 
The time reversal operator is antiunitary. It reverses the momentum of a particle and flips its spin. Effectively, 
$\mathcal{O}(\vec{x},t) \; \stackrel{\mathcal{T}}{\longrightarrow} \; \mathcal{O} (\vec{x},-t)$ and 
$\alpha \stackrel{\mathcal{T}}{\longrightarrow} \; \alpha^{\ast}$. 
In the weak eigenstates, the interactions $\mathcal{L}_{cc}$ in Eq.~\ref{eq:Lcc} are invariant under $\mathcal{CP}$ and $\mathcal{T}$, as all coupling constants are real. On the other hand, the Yukawa interactions violate both $\mathcal{CP}$ and $\mathcal{T}$. Using the up-quark sector again as an example, for each conjugate pair specified by indices $i$ and $j$,
\begin{eqnarray}
\overline{U}_{R,i} (M_{u})_{ij} Q_{L,j}  +  \overline{Q}_{L,j} (M_{u}^{\dagger})_{ji} U_{R,i}  
\stackrel{\mathcal{CP}}{\longrightarrow} 
 \overline{Q}_{L,j} (M_{u})_{ij} U_{R,i}   +   \overline{U}_{R,i} (M_{u})_{ij}^{\ast} Q_{L,j} \; , \\
\overline{U}_{R,i} (M_{u})_{ij} Q_{L,j} +  \overline{Q}_{L,j} (M_{u}^{\dagger})_{ji} U_{R,i} 
 \stackrel{\mathcal{T}}{\longrightarrow} 
 \overline{U}_{R,i} (M_{u})_{ij}^{\ast} Q_{L,j}  +  \overline{Q}_{L,j} (M_{u})_{ij} U_{R,i} \; .
\end{eqnarray}
The complexity of the mass matrix, giving rise to $\mathcal{CP}$ and $\mathcal{T}$ violations, ensues from  the complex CG coefficients in $T^{\prime}$. 
Here we have suppressed the space-time coordinates, the inversions of which under the transformations are assumed implicitly. Due to its hermiticity, the Lagrangian is $\mathcal{CPT}$ invariant,
\begin{equation}
\overline{U}_{R} M_{u} Q_{L} + \overline{Q}_{L} M_{u}^{\dagger} U_{R}  \stackrel{\mathcal{CPT}}{\longrightarrow} 
 \overline{Q}_{L} M_{u}^{\dagger} U_{R} + \overline{U}_{R} M_{u} Q_{L} \; .
 \end{equation}
Alternatively, in the mass eigenstates, the Yukawa interactions are invariant under $\mathcal{CP}$ and $\mathcal{T}$, while the charged current interactions violate $\mathcal{CP}$ and $\mathcal{T}$ individually and are invariant under $\mathcal{CPT}$. Note that CP violation is inherent in the Lagrangian~Eq.~\ref{eq:Ltt}-\ref{eq:Lff}, which is $T^{\prime}$ and $SU(5)$ invariant.

{\it Numerical Predictions.}---The predicted charged fermion mass matrices in our model are parametrized in terms of 7 parameters~\cite{Chen:2007afa},
\begin{eqnarray}
\frac{M_{u}}{y_{t} v_{u}} & = & \left( \begin{array}{ccccc}
i g & ~~ &  \frac{1-i}{2}  g & ~~ & 0\\
\frac{1-i}{2} g & & g + (1-\frac{i}{2}) h  & & k\\
0 & & k & & 1
\end{array}\right)  , \\
\frac{M_{d}, \; M_{e}^{T}}{y_{b} v_{d} \phi_{0}\zeta_{0}}  & = &  \left( \begin{array}{ccccc}
0 & ~~ & (1+i) b & ~~ & 0\\
-(1-i) b & & (1,-3) c & & 0\\
b & &b & & 1
\end{array}\right)  \; .
\end{eqnarray}
With $b \equiv \phi_{0} \psi^{\prime}_{0} /\zeta_{0} = 0.00304$, $c\equiv \psi_{0}N_{0}/\zeta_{0}=-0.0172$,  $k \equiv y^{\prime}\psi_{0}\zeta_{0}=-0.0266$, $h\equiv \phi_{0}^{2}=0.00426$ and $g \equiv \phi_{0}^{\prime 3}= 1.45\times 10^{-5}$, the following mass ratios are obtained, 
$m_{d}: m_{s} : m_{b} \simeq \theta_{c}^{\scriptscriptstyle 4.7} : \theta_{c}^{\scriptscriptstyle 2.7} : 1$, 
$m_{u} : m_{c} : m_{t} \simeq  \theta_{c}^{\scriptscriptstyle 7.5} : \theta_{c}^{\scriptscriptstyle 3.7} : 1$, 
with $\theta_{c} \simeq \sqrt{m_{d}/m_{s}} \simeq 0.225$. (These ratios in terms of $\theta_{c}$ coincide with those give in ~\cite{McKeen:2007ry}.) We have also taken $y_{t} = 1.25$ and $y_{b}\phi_{0} \zeta_{0} \simeq m_{b}/m_{t} \simeq 0.011$ and have taken into account the renormalization group corrections. As a result  of the GJ relations, realistic charged lepton masses are obtained. 
Making use of these parameters, the complex CKM matrix is,
\begin{eqnarray}
\left( \begin{array}{ccc}
0.974e^{-i 25.4^{o}} & 0.227 e^{i23.1^{o}} & 0.00412e^{i166^{o}} \\
0.227 e^{i123^{o}} & 0.973 e^{-i8.24^{o}} & 0.0412 e^{i180^{o}} \\
0.00718 e^{i99.7^{o}} & 0.0408 e^{-i7.28^{o}} & 0.999
\end{array}\right). 
\end{eqnarray}
Values for all $|V_{\mbox{\tiny CKM}}|$ elements are consistent with current experimental values~\cite{Amsler:2008zzb} except for $|V_{td}|$, the experimental determination of which has large hadronic uncertainty.  Further, as so far no $t \rightarrow d$ transition at tree level has been observed, the experimental extraction of $|V_{td}|$ is based on loop calculations done in the framework of the Standard Model. 

The predictions of our model for the angles in the unitarity triangle are, 
$\beta = 23.6^{o}$ ($\sin2\beta  =  0.734$), $\alpha = 110^{o}$, and $\gamma = \delta_{q} = 45.6^{o}$, 
(where $\delta_{q}$ is the CP phase in the standard parametrization), and they agree with the direct measurements within $1\sigma$ of BaBar and $2\sigma$ of Belle (M. Antonelli et al in Ref.~\cite{Amsler:2008zzb}.) 
Except for observables whose experimental values are obtained from direct measurements, comparison between the global fit results and predictions of new physics models is {\it not} appropriate, because the global fit is based on the Standard Model with loop corrections. (Nevertheless, even in this case, our predictions for the Wolfenstein paramteres, $\lambda=0.227$, $A=0.798$, $\overline{\rho} = 0.299$ and $\overline{\eta}=0.306$, are very close to the global fit values except for $\overline{\rho}$. Our prediction for the Jarlskog invariant, $J  \equiv  \mbox{Im} (V_{ud} V_{cb} V_{ub}^{\ast} V_{cd}^{\ast}) = 2.69 \times 10^{-5}$, in the quark sector also agrees with the current global fit value.) Potential direct measurements for these parameters at the LHCb can test our predictions. 

As a result of the GJ relations, our model predicts the sum rule~\cite{Chen:2007afa,Antusch:2005kw} between the solar neutrino mixing angle and the Cabibbo angle in the quark sector, $\tan^{2} \theta_{\odot} \simeq \tan^{2} \theta_{\odot,\mbox{\tiny TBM}} + \frac{1}{2} \theta_{c} \cos\delta_{\ell}$, with $\delta_{\ell}$ being the leptonic Dirac CP phase in the standard parametrization.  
In addition, our model predicts $\theta_{13} \sim \theta_{c}/3\sqrt{2}$. Numerically, the diagonalization matrix for the charged lepton mass matrix combined with $U_{TBM}$ gives the PMNS matrix, 
\begin{equation}
\left( \begin{array}{ccc}
0.838 e^{-i178^{o}} & 0.543 e^{-i173^{o}} & 0.0582 e^{i138^{o}}  \\
0.362 e^{-i3.99^{o}} & 0.610 e^{-i173^{o}} & 0.705 e^{i3.55^{o}}  \\
0.408 e^{i180^{o}} & 0.577 & 0.707
\end{array}\right) \; ,
\end{equation}
which gives $\sin^{2}\theta_{\mathrm{atm}} = 1$, $\tan^{2}\theta_{\odot} = 0.420$ and $|U_{e3}| = 0.0583$. The two VEV's, $u_{0} = -0.0593$ and $\xi_{0} = 0.0369$, give $\Delta m_{atm}^{2} = 2.4 \times 10^{-3} \; \mbox{eV}^{2}$ and $\Delta m_{\odot}^{2} = 8.0 \times 10^{-5} \; \mbox{eV}^{2}$. As the three masses are given in terms of two VEV's, there exists a mass sum rule, 
$m_{1} - m_{3} = 2m_{2}$, leading to normal mass hierarchy, $\Delta m_{atm}^{2} > 0$~\cite{Chen:2007afa}.   
The leptonic Jarlskog is predicted to be $J_{\ell} = -0.00967$, and equivalently, this gives  a Dirac CP phase, $\delta_{\ell} = 227^{o}$. With such $\delta_{\ell}$, the correction from the charged lepton sector can account for the difference between the TBM prediction and  the current best fit value for $\theta_{\odot}$. Our model predicts $(m_{1},m_{2},m_{3}) = (0.0156,-0.0179,0.0514)$ eV, with Majorana phases $\alpha_{21} = \pi$ and $\alpha_{31}$ = 0.  

Our model has nine input parameters, predicting a total of twenty-two physical quantities: 12 masses, 6 mixing angles, 2 Dirac CP violating phases and 2 Majorana phases. Our model is testable by more precise experimental values for $\theta_{13}$, $\tan^{2}\theta_{\odot}$ and $\gamma$ in the near future. 
$\delta_{\ell}$ is the only non-vanishing leptonic CP violating phase in our model and it gives rise to lepton number asymmetry, $\epsilon_{\ell} \sim 10^{-6}$. By virtue of leptogenesis, this gives the right sign and magnitude of the matter-antimatter asymmetry~\cite{CM}.

{\it Conclusion.}---We propose the complex group theoretical CG coefficients as a novel origin of CP violation. This is manifest in our model based on SU(5)  combined with the double tetrahedral group, $T^{\prime}$. Due to the presence of the doublet representations in $T^{\prime}$, there exist complex CG coefficients, leading to explicit CP violation in the model, while having real Yukawa couplings and  scalar VEVs. The predicted CP violation measures in the quark sector are consistent with the current experimental data. The leptonic Dirac CP violating phase is predicted to be $\delta_{\ell} \sim 227^{o}$, which gives the cosmological matter asymmetry.  


\begin{acknowledgments}
The work of M-CC was supported, in part, by the National Science Foundation under grant no. PHY-0709742. The work of KTM was supported, in part, by the Department of Energy under Grant no. DE-FG02-04ER41290. 
\end{acknowledgments}

\end{document}